\documentclass[twocolumn]{aastex701}

\usepackage{bm}
\usepackage{amsmath}

\begin{document}

\title{Investigating tidal stripping of a pre-existing moon as the origin of Saturn's young icy rings}

\author[orcid=0000-0003-1097-0521]{Yifei Jiao}
\affiliation{Department of Earth and Planetary Sciences, University of California, Santa Cruz, CA 95064, USA}
\affiliation{Tsinghua University, Beijing 100084, China}
\email[show]{jiaoyf.thu@gmail.com}

\author[orcid=0000-0003-3573-5915]{Francis Nimmo}
\affiliation{Department of Earth and Planetary Sciences, University of California, Santa Cruz, CA 95064, USA}
\email[]{}

\author[orcid=0000-0002-6142-0653]{Jack Wisdom}
\affiliation{Department of Earth, Atmospheric, and Planetary Sciences, Massachusetts Institute of Technology, Cambridge, MA 02139, USA}
\email[]{}

\author[orcid=0000-0003-4423-1647]{Rola Dbouk}
\affiliation{Department of Earth, Atmospheric, and Planetary Sciences, Massachusetts Institute of Technology, Cambridge, MA 02139, USA}
\affiliation{Department of Earth, Environmental, and Planetary Sciences, Brown University, Providence, RI 02912, USA}
\email[]{}

\begin{abstract}

The origin of Saturn's rings has been debated for decades. Measurements from Voyager and Cassini have suggested that the rings could be as young as $\sim100$~Myr and composed of nearly pure water ice.
Several scenarios have been proposed to explain these properties. 
One hypothesis \citep{wisdom2022loss} is that the rings formed through the recent tidal disruption of a pre-existing moon, Chrysalis, which experienced a close encounter with Saturn following its highly eccentric orbit.
However, the mechanism by which this hypothesis would have formed the rings remains largely unexplored, in particular, whether Chrysalis could supply ring material of the desired mass and composition.
To address these questions, we perform smoothed particle hydrodynamics simulations to investigate the tidal response of Chrysalis during close encounters with Saturn.
Our results demonstrate that preferential tidal stripping of the ice mantle from a differentiated Chrysalis can produce rings with both mass and composition resembling the present rings---provided that the closest encounter occurs between the parabolic Roche limits for ice $\sim1.53R_{\rm S}$ and rock $\sim1.07R_{\rm S}$---consistent with \cite{wisdom2022loss}.
Moreover, multiple close encounters can extend the effective disruption limit by spinning up the body, enhancing the tidal stripping efficiency.
Following close encounters, the rocky remnant of Chrysalis would have been removed in less than few kyr, either by collision with Saturn or ejection onto a hyperbolic orbit.
These findings support the hypothesis that Saturn's rings could originate from a recent lost moon, and imply a highly dynamical evolution of the Saturnian system over the past few hundred million years.

\end{abstract}



\section{Introduction}

The Saturnian system, with its prominent rings and diverse moons, has posed longstanding questions regarding its formation and evolution since the first telescopic observations in the 17th century.
While the large moons are generally (though not universally) thought of as primordial, having formed during the early stages of Saturn's history \citep{blanc2025understanding}, the origin of the rings---which may be young \citep{zhang2017exposure}---remains uncertain and has been debated for decades.
The rings of Saturn occupy the region from just above planet's surface to approximately 2.5 Saturn radii ($R_{\rm S}$), which corresponds to the classical Roche limit \citep{roche1847memoire}.
The rings are composed of $\gtrsim95$~wt.\% water ice \citep{doyle1989radiative,cuzzi1998compositional}, with a total mass of $(1.54\pm0.49)\times10^{19}$~kg, corresponding to $(0.41\pm0.13)M_{\rm Mimas}$ \citep{iess2019measurement}, and a particle size range from centimeters to meters \citep{cuzzi2009ring}.
These physical signatures serve as a vital constraint for ring age estimates, which can be derived from (1) exposure models depending on the interplanetary micrometeoroid impact flux and pollution efficiency \citep{zhang2017exposure}, and (2) dynamical evolution models including viscous spreading \citep{crida2019saturn} and micrometeoroid-driven processes such as ballistic transportation and mass loading \citep{estrada2023constraints}.
Together, these analyses consistently indicate that Saturn's rings are likely only a few hundred million years old.

Various models have been proposed to explain the origin of Saturn's rings, which can be broadly grouped into two mechanisms: collisional breakup or tidal disruption of a parent body.
Collisional origin scenarios include (1) a mid-size inner moon being disrupted by a heliocentric impactor, which would favor older rings \citep{teodoro2023recent}; and (2) mutual collisions between two mid-size moons, which suggest a young age for both the rings and some inner moons \citep{cuk2016dynamical}.
Tidal disruption scenarios require a body to approach Saturn closely enough to experience strong tidal forces. These scenarios include (1) a Titan-size inner moon being tidally stripped of its ice layer when migrating into the Roche limit, leaving an ancient, massive ring \citep{canup2010origin}; (2) a large passing comet being tidally disrupted, with some ice materials captured to form the rings, also favoring old rings \citep{dones1991recent,hyodo2017ring}; and (3) a pre-existing moon being destabilized to have a close encounter with Saturn and be tidally disrupted \citep{wisdom2022loss}. This model is consistent with a young ring age but did not investigate the ring mass or composition.

Here we focus on the recent-lost-moon hypothesis proposed by \cite{wisdom2022loss}, which offers a promising explanation for Saturn's rings origin.
In this scenario, Saturn once hosted an additional satellite, Chrysalis, and Saturn was in spin-orbit precession resonance with Neptune. The fast outward migration of Titan would have increased Saturn's obliquity through this resonance \citep{saillenfest2021large}. 
Chrysalis was then caught in a 3:1 mean motion resonance with Titan, became destabilized, and was eventually removed $\sim100$~Myr ago, allowing Saturn to escape the resonance with Neptune.
This process can explain Titan's orbital eccentricity, as required from its highly dissipative interior \citep{petricca2025titan}.
Before its loss, Chrysalis is hypothesized to have undergone a close encounter with Saturn, causing tidal disruption and forming the rings \citep{wisdom2022loss}.
However, there are two major concerns regarding this hypothesis \citep{crida2025age}: (1) the present ring mass is only about 1\% of the mass of Chrysalis, while the ring mass can not decrease significantly within a few hundred million years once circularized; (2) Chrysalis is argued to have had a small periapsis of $q<1.2R_{\rm S}$ to match the specific angular momentum of the rings. At this distance, rocky material would be disrupted, which conflicts with the rings' ice-rich composition.

Although previous tidal disruption scenarios have yielded useful insights to help address these concerns, they are not directly applicable to the Chrysalis scenario.
For example, \cite{canup2010origin} assumed a Titan-size satellite on a nearly circular orbit at the Roche limit, whereas Chrysalis has a highly eccentric orbit and is less massive. \cite{hyodo2017ring} considered the tidal disruption of a hyperbolic object and focused exclusively on the captured debris, while Chrysalis was initially a bound satellite.
These considerations motivate further investigation of the tidal response of Chrysalis to test the recent-lost-moon hypothesis.

In this work, we perform smoothed particle hydrodynamics (SPH) simulations to investigate the tidal response of Chrysalis during close encounters with Saturn, based on the physical and orbital constraints established by \cite{wisdom2022loss}.
We explore a parameter space defined by the body's ice-to-rock ratio, the pre-encounter periapsis distance $q_0$, and the spin rate change during multiple encounters. These simulations allow us to characterize the resulting rings in terms of mass, composition, specific angular momentum, and dynamical evolution, enabling a direct test of the recent-lost-moon hypothesis.

\section{Parameter Space}\label{sec2}

\subsection{Physical constraints}\label{sec2.1}

\begin{figure}[b!]
\centering
\includegraphics[width=\linewidth]{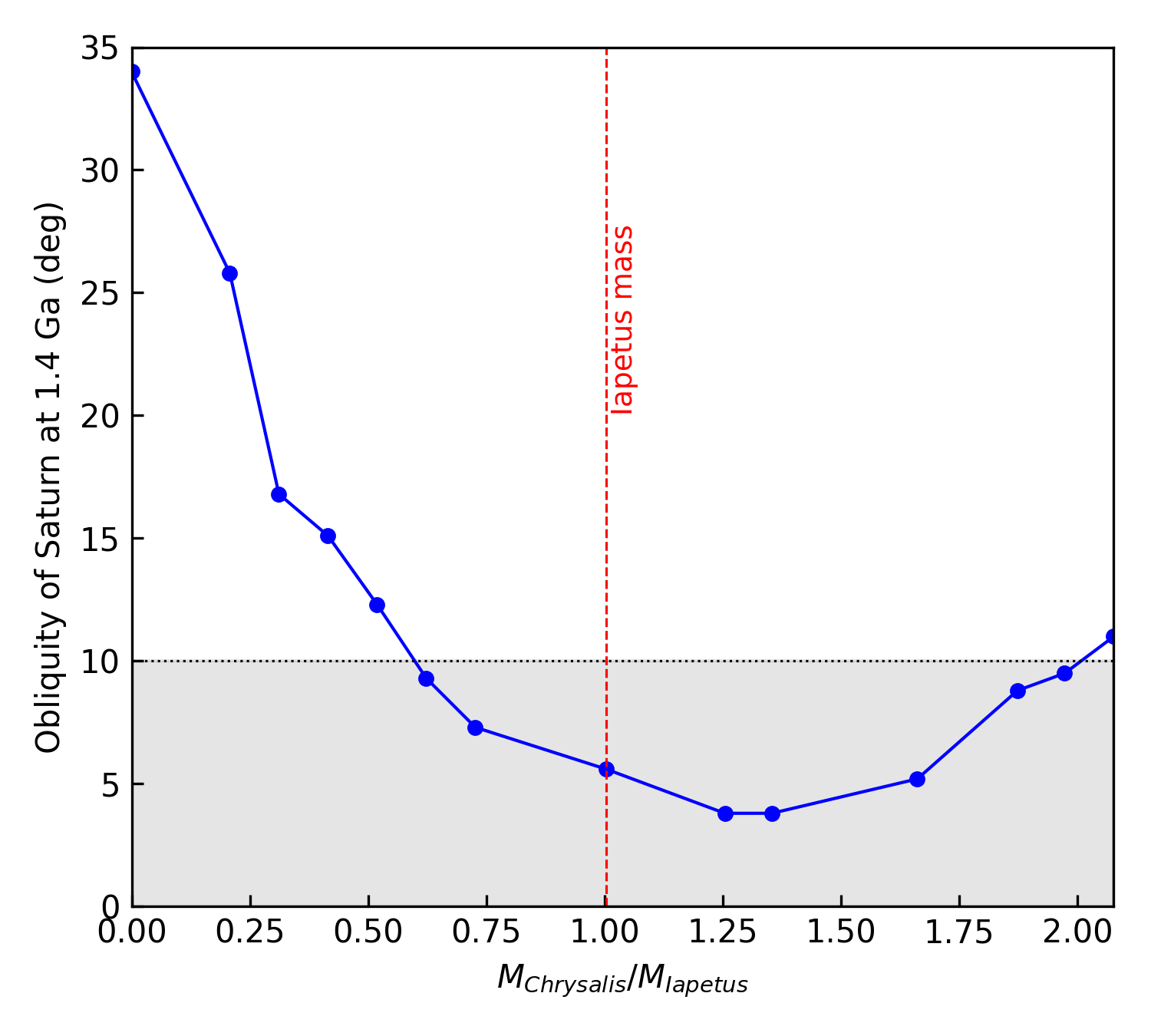}
\caption{Mass constraint of Chrysalis according to \cite{wisdom2022loss}. Each point represents a simulation of the resonance model backward in time over 1.4 Gyr, for given Chrysalis mass. The final obliquity should be below about $10^{\circ}$, i.e., the shadow area, requiring the mass of Chrysalis from about $0.6M_{\rm Iapetus}$ to $2.0M_{\rm Iapetus}$.}
\label{fig:chrysalis-mass}
\end{figure}

The mass of Chrysalis is constrained by the requirement that the precession constant has the required change, so that Saturn can exit the resonance with Neptune, leaving a large obliquity as seen today.
Figure~\ref{fig:chrysalis-mass} shows how it works: using the resonance model in \cite{wisdom2022loss}, the mass of Chrysalis must be within 0.6--2.0$M_{\rm Iapetus}$ to match a primordially low obliquity of Saturn.
A satellite with this mass is typically differentiated due to radioactive decay and/or tidal heating \citep{neveu2019evolution}.
However, the ice-to-rock ratio of Chrysalis cannot be directly constrained from dynamical models. As a reference, other Saturnian moons of comparable mass have bulk densities from about 1~g\,cm$^{-3}$ to 1.5~g\,cm$^{-3}$ \citep{nimmo2016ocean}. Assuming an ice density of 0.92~g\,cm$^{-3}$ and a rock density of 3~g\,cm$^{-3}$, these values imply an ice mass fraction of about 90\%--50\%.
The pre-encounter spin of Chrysalis is also unknown. Before its destabilization, it likely had a synchronous spin period of about 48 days, corresponding to a 3:1 resonance with Titan. This spin state could have been further altered by tidal torques during its close encounter with other moons, such as Titan, prior to Chrysalis's approach to Saturn. However, these processes are out of the scope of this work.
We therefore assume a fixed mass of $\sim M_{\rm Iapetus}$ and initially no spin for Chrysalis. To model the differentiated interior, we adopt two representative ice mass fractions of 50~wt.\% (Dione-like) and 80~wt.\% (Iapetus-like) in our simulations.

\subsection{Pre-encounter orbits}\label{sec2.2}

\begin{figure*}[htb!]
\centering
\includegraphics[width=\linewidth]{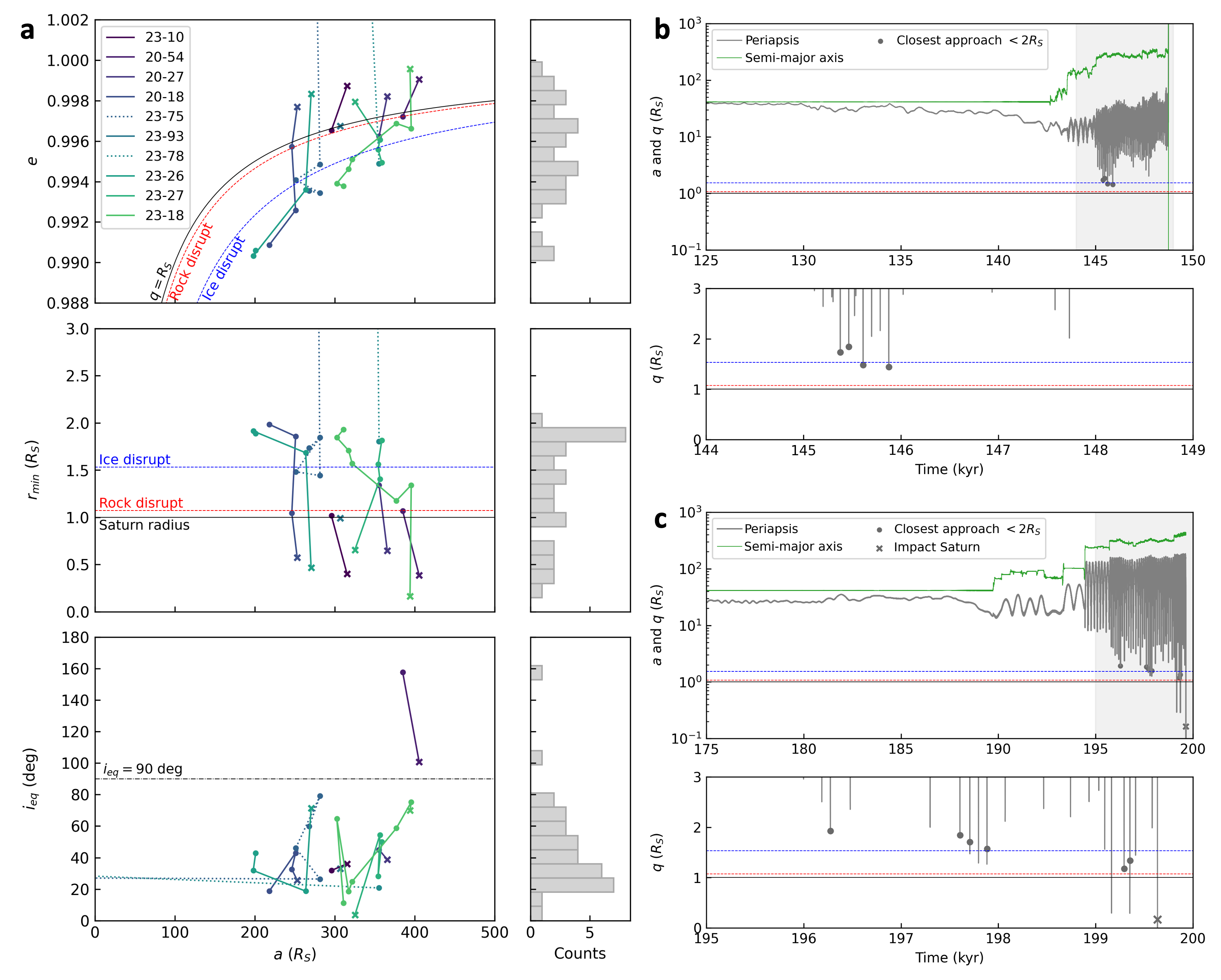}
\caption{Saturn-grazing cases from \cite{wisdom2022loss} with closest distance $<2R_{\rm S}$. (a) Orbit elements of Chrysalis at each closest approach, where the solid-line cases end with impacting Saturn (cross marked), and the dotted-line cases ultimately become hyperbolic. The parabolic Roche limits \citep{sridhar1992tidal} for ice and rock are provided for reference. (b) Orbit evolution of case 23-75 from panel (a), with the shadow region enlarged to show the close encounter details. This case ends in a hyperbolic orbit with a negative semi-major axis. (c) Same as panel (b) but for case 23-18, which ends with impact on Saturn. Note that a close approach (dot points) requires a small transient periapsis $q$, but the reverse does not hold: a small $q$ does not guarantee a close approach.}
\label{fig:para-orbit}
\end{figure*}

Regarding the orbital parameters, Figure~\ref{fig:para-orbit} shows ten Saturn-grazing cases from \cite{wisdom2022loss}, defined as encounters whose initial closest-approach distance to Saturn is less than $2.5R_{\rm S}$, where $R_{\rm S}$ is the Saturn radius.
In all cases, Chrysalis subsequently experienced one or more encounter with $q<2R_{\rm S}$ before colliding with Saturn (80\%) or escaping the system (20\%).
The grazing orbits are characterized with semi-major axes $a$ of about $200R_{\rm S}$--$400R_{\rm S}$ and high eccentricities, exceeding $0.99$.
The orbital inclinations relative to Saturn's equatorial plane, $i_{\rm eq}$, are generally prograde ($<90^\circ$, though with one retrograde exception), consistent with the orientation of the present rings.
In most cases, Chrysalis would have an inclination exceeding $\sim20^{\circ}$, preventing encounters with the inner moons---one of the concerns raised by \cite{crida2025age}.

The outcome of a close encounter is primarily determined by the pericenter distance and the duration of tidal forcing.
Since Chrysalis has a highly eccentric orbit, we introduce the parabolic Roche limit \citep{sridhar1992tidal} as
\begin{equation}
    R_{\rm roche} = 1.69R_{\rm S}\left (\frac{\rho_{\rm S}}{\rho_0}\right )^\frac{1}{3},
\end{equation}
where $\rho_{\rm S}=0.687$~kg\,m$^{-3}$ is Saturn's mean density, and $\rho_0$ is the density of the passing body.
For icy material, this yields a tidal disruption distance of $\sim1.53R_{\rm S}$, which lies within the classical circular-orbit Roche limit $\sim2.5R_{\rm S}$. For rock material, the parabolic Roche limit is only $\sim1.07R_{\rm S}$, which is just above Saturn's surface.
A differentiated body composed of an ice mantle and rock core, such as Chrysalis, would therefore have a disruption distance between these two limits.
Since $q \ll a$, the velocity at periapsis, $v_q = \sqrt{GM_{\rm S}(2/q-1/a)}$, is dominated by the $2/q$ term and is largely insensitive to the semi-major axis. For our modeling below, we neglect the variations in semi-major axis and adopt a fixed value $a_0 = 200R_{\rm S}$, while varying the periapsis distance $q_0$ from $R_{\rm S}$ to $\sim 2R_{\rm S}$.
In our simulations, inclination is not explicitly modeled, as Saturn is treated as a central point mass. Under this assumption, inclined configurations can be obtained by a simple coordinate rotation.
For more realistic scenarios involving multiple passages, e.g., Figure~\ref{fig:para-orbit}~(b–c), we adopt the specific $(a_0, q_0)$ values from the dynamical simulation results of \cite{wisdom2022loss}.

The final parameters used in our simulations are summarized in Table~\ref{tab:simu_setup}.
Generally, we consider four sets of cases: d-cases (Dione-like body, 50~wt.\% ice, varying periapsis distances), i-cases (Iapetus-like body, 80~wt.\% ice, varying periapsis distances), h-cases (multiple encounters ending in a hyperbolic escape), and s-cases (multiple encounters ending in a Saturn impact).

\begin{deluxetable*}{crrrrrrrl}[htb!]
\tabletypesize{\scriptsize}
\tablecaption{SPH simulations setup. The d-cases vary the periapsis $q_0$, with a fixed semi-major axis of $a_0=200~R_{\rm S}$, no spin, and an ice mass fraction of 50\% (Dione-like). Case d6 is a high-resolution rerun  of case d5, with $2\times10^5$ particles. For comparison, the i-cases adopt a higher ice mass fraction of 80\% (Iapetus-like). The h-cases reproduce the four close encounters before Chrysalis becomes hyperbolic, while the s-cases reproduce the six encounters before a Saturn impact, where the core would be disrupted during the fifth encounter and then the simulation is terminated. In both cases, the specific $q_0$ and $a_0$ are taken from dynamical simulations of \cite{wisdom2022loss}, as shown in Figure~\ref{fig:para-orbit}~(b--c). Also note that the orbital inclination has been ignored for simplification, leading to an overestimate of the spin change during multiple encounters.}
\label{tab:simu_setup}
\tablewidth{0pt} 
\tablehead{
\colhead{Case no.} & \colhead{$M_0$ (kg)} & \colhead{Ice (wt.\%)} & \colhead{$q_0$ ($R_{\rm S}$)} & $a_0$ ($R_{\rm S}$) & \colhead{$T_{\rm spin,0}$ (h)} & $M_{\rm bound}/M_{\rm ring}$ & Bound ice (wt.\%) & Notes
}
\startdata
d1 & $2.01\times10^{21}$ & 50.0 & 1.00 & 200.0 & $\infty$ & 0.04 & 89.7 & intersect with Saturn \\
d2 & $2.01\times10^{21}$ & 50.0 & 1.05 & 200.0 & $\infty$ & 78.27 & 46.1 & core disrupted \\
d3 & $2.01\times10^{21}$ & 50.0 & 1.10 & 200.0 & $\infty$ & 14.53 & 91.2 & tidal stripping \\
d4 & $2.01\times10^{21}$ & 50.0 & 1.20 & 200.0 & $\infty$ & 12.60 & 98.5 & tidal stripping \\
d5 & $2.01\times10^{21}$ & 50.0 & 1.30 & 200.0 & $\infty$ & 5.36 & 100.0 & tidal stripping \\
d6 & $2.01\times10^{21}$ & 50.0 & 1.30 & 200.0 & $\infty$ & 6.50 & 100.0 & (high resolution) \\
d7 & $2.01\times10^{21}$ & 50.0 & 1.40 & 200.0 & $\infty$ & 1.86 & 100.0 & tidal stripping \\
d8 & $2.01\times10^{21}$ & 50.0 & 1.45 & 200.0 & $\infty$ & 0.93 & 100.0 & tidal stripping \\
d9 & $2.01\times10^{21}$ & 50.0 & 1.50 & 200.0 & $\infty$ & - & - & spin up \\
d10 & $2.01\times10^{21}$ & 50.0 & 1.80 & 200.0 & $\infty$ & - & - & spin up \\
d11 & $2.01\times10^{21}$ & 50.0 & 2.10 & 200.0 & $\infty$ & - & - & spin up \\
\hline
i12 & $2.01\times10^{21}$ & 80.0 & 1.00 & 200.0 & $\infty$ & 0.20 & 94.9 & intersect with Saturn \\
i13 & $2.01\times10^{21}$ & 80.0 & 1.05 & 200.0 & $\infty$ & 21.72 & 98.8 & tidal stripping \\
i14 & $2.01\times10^{21}$ & 80.0 & 1.10 & 200.0 & $\infty$ & 29.28 & 99.8 & tidal stripping \\
i15 & $2.01\times10^{21}$ & 80.0 & 1.20 & 200.0 & $\infty$ & 25.46 & 100.0 & tidal stripping \\
i16 & $2.01\times10^{21}$ & 80.0 & 1.30 & 200.0 & $\infty$ & 20.15 & 100.0 & tidal stripping \\
i17 & $2.01\times10^{21}$ & 80.0 & 1.40 & 200.0 & $\infty$ & 12.77 & 99.9 & tidal stripping \\
i18 & $2.01\times10^{21}$ & 80.0 & 1.50 & 200.0 & $\infty$ & 4.98 & 100.0 & tidal stripping \\
i19 & $2.01\times10^{21}$ & 80.0 & 1.60 & 200.0 & $\infty$ & 1.30 & 100.0 & tidal stripping \\
\hline
h20 & $2.01\times10^{21}$ & 50.0 & 1.734 & 268.1 & $\infty$ & - & - & spin up \\
h21 & $2.01\times10^{21}$ & 50.0 & 1.846 & 281.5 & 20.84 & - & - & spin up \\
h22 & $2.01\times10^{21}$ & 50.0 & 1.482 & 251.2 & 14.87 & 1.21 & 100.0 & tidal stripping \\
h23 & $1.98\times10^{21}$ & 49.4 & 1.442 & 281.4 & 5.68 & 9.11 & 100.0 & tidal stripping \\
\hline
s24 & $2.01\times10^{21}$ & 50.0 & 1.930 & 310.9 & $\infty$ & - & - & spin up \\
s25 & $2.01\times10^{21}$ & 50.0 & 1.845 & 302.5 & 52.37 & - & - & spin up \\
s26 & $2.01\times10^{21}$ & 50.0 & 1.707 & 317.3 & 23.50 & - & - & spin up \\
s27 & $2.01\times10^{21}$ & 50.0 & 1.569 & 321.7 & 11.76 & - & - & spin up \\
s28 & $2.01\times10^{21}$ & 50.0 & 1.176 & 377.0 & 5.96 & 52.81 & 54.1 & core disrupted \\
\enddata
\end{deluxetable*}

\section{From tidal stripping to rings}\label{sec3}

\subsection{Smoothed particle hydrodynamics}\label{sec3.1}

With the parameter space defined in Section~\ref{sec2}, we perform smoothed particle hydrodynamics (SPH) simulations to investigate the tidal disruption/stripping of Chrysalis.
SPH is a mesh-free Lagrangian particle method that represents physical quantities using a smoothing kernel and solves the hydrodynamic conservation equations for mass, momentum, and energy \citep{monaghan1992smoothed}.
The SPH method has been widely used to model hypervelocity impacts and tidal responses of planetary bodies, e.g., \cite{benz1994impact,hyodo2017ring}.
In this work, we use the SPH code adapted from \cite{jiao2024sph,jiao2024asteroid}, which has been extensively validated to ensure its numerical accuracy and performance (see Appendix).
Since Chrysalis could be as massive as Iapetus and likely differentiated, we model it as a mantle-core-layered body with initial density and pressure profiles in hydrostatic equilibrium \citep{ruiz2021effect}.
The mantle is assumed pure water ice and the core is silicate, described using the Tillotson equation of states of ice and basalt, respectively \citep{melosh1989impact}.
Self-gravity is approximated using the Barnes-Hut algorithm \citep{barnes1986hierarchical}; tidal forces are computed according to Eq.~(\ref{eq:tide}).
The total number of SPH particles is $1\times10^5$ in most of the cases, corresponding to a particle scale of $\sim$25~km.
To evaluate numerical convergence, we also perform an additional high-resolution run with $2\times10^5$ particles (case d6).
All simulations start 4~h before the body's closest approach with Saturn (outside the classical Roche limit) and run for 50~h, by which time the tidally induced debris clusters reach a stable distribution.
The system is integrated using a second-order leapfrog scheme with an adaptive timestep of order 1~s, which is determined by the Courant-Friedrichs-Lewy (CFL) condition \citep{courant1928partiellen}.

\subsection{Tidal stripping of Chrysalis}\label{sec3.2}

\begin{figure*}[htb!]
\centering
\includegraphics[width=0.95\linewidth]{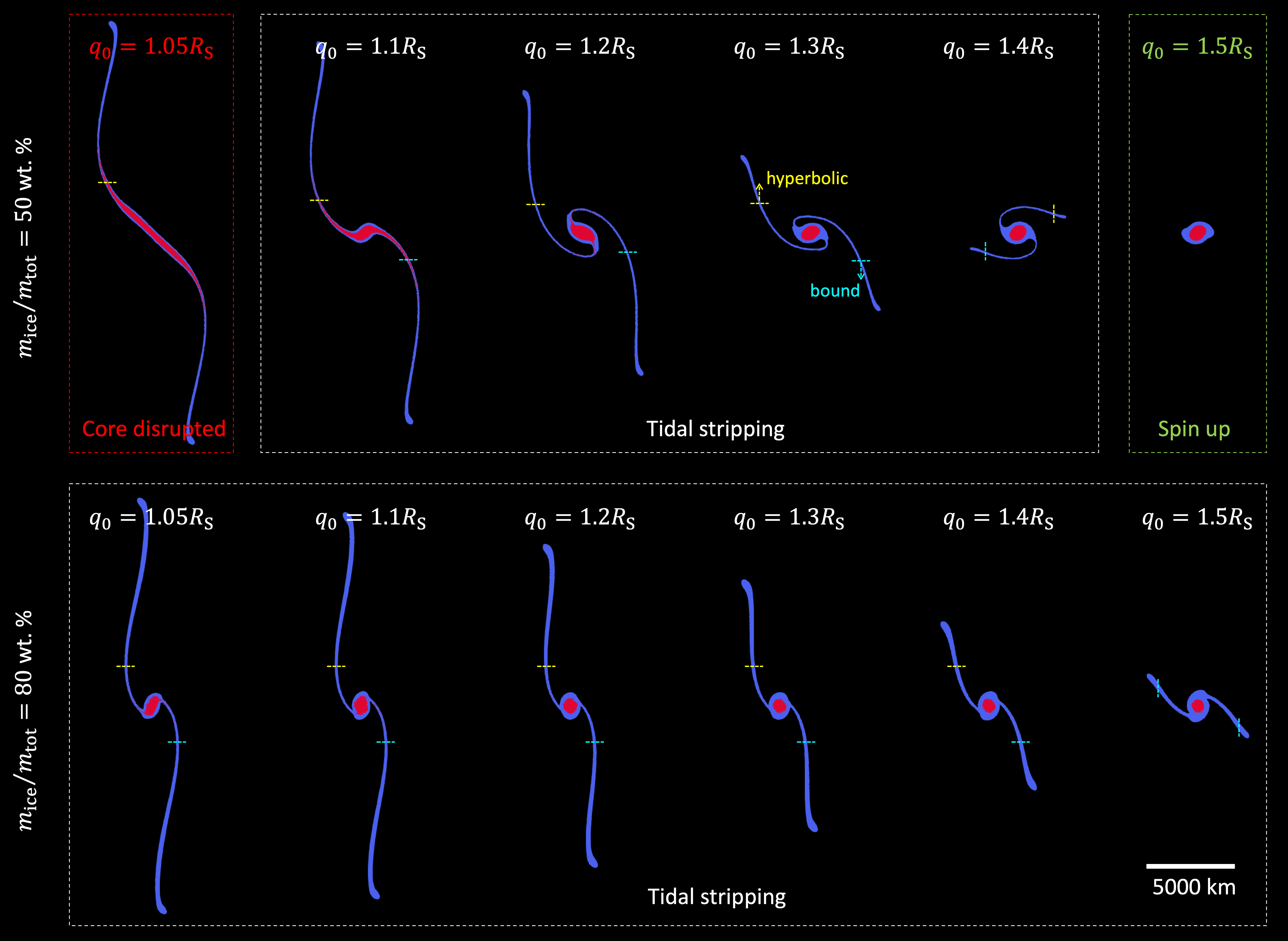}
\caption{SPH simulation snapshots with varying ice-to-rock ratios and periapses (d- and i- cases in Table~\ref{tab:simu_setup}). Red and blue particles represent the rock core and ice mantle, respectively. In all cases, the snapshot corresponds to 6~h after periapsis, not the simulation end time. The dashed lines indicate the final fates of particles, with yellow for hyperbolic and cyan for bound (excluding the largest remnant).}
\label{fig:snapshot-all}
\end{figure*}

Our simulation results suggest various tidal responses of Chrysalis, primarily determined by its periapsis, as shown in Figure~\ref{fig:snapshot-all}.
The results can be classified into four regimes: (1) intersect with Saturn when $q_0$ is close to $R_{\rm S}$, (2) core disruption when $q_0$ is close to the rock disruption limit, (3) tidal stripping when $q_0$ lies between the ice and rock disruption limits, and (4) spin up when $q_0$ is beyond the ice disruption limit. Of these possibilities, the tidal stripping regime would be preferred to form an ice-rich ring.
The exact boundaries between regimes are also affected by the ice-to-rock ratios (see also Figure~\ref{fig:snapshot-all}).
A higher rock fraction (d-cases, 50~wt.\% ice) increases the bulk density and thus self gravity, making the body more resistant to tidal stripping at a given $q_0$, but leading to more catastrophic destruction once the rock disruption limit is crossed.
Conversely, a higher ice fraction (i-cases, 80~wt.\% ice) lowers the bulk density and widens the range of $q_0$ over which partial tidal stripping occurs.
For example, at $q_0 = 1.05 R_{\rm S}$---within the rock disruption limit---the ice mantle is strongly stripped while the small rocky core remains almost intact for 80~wt.\% ice. At $q_0 = 1.5 R_{\rm S}$, where tidal forces are weaker, some ice can still be stripped off (80~wt.\% ice) rather than merely inducing rotational spin-up (50~wt.\% ice).

\begin{figure*}[htb!]
\centering
\includegraphics[width=0.94\linewidth]{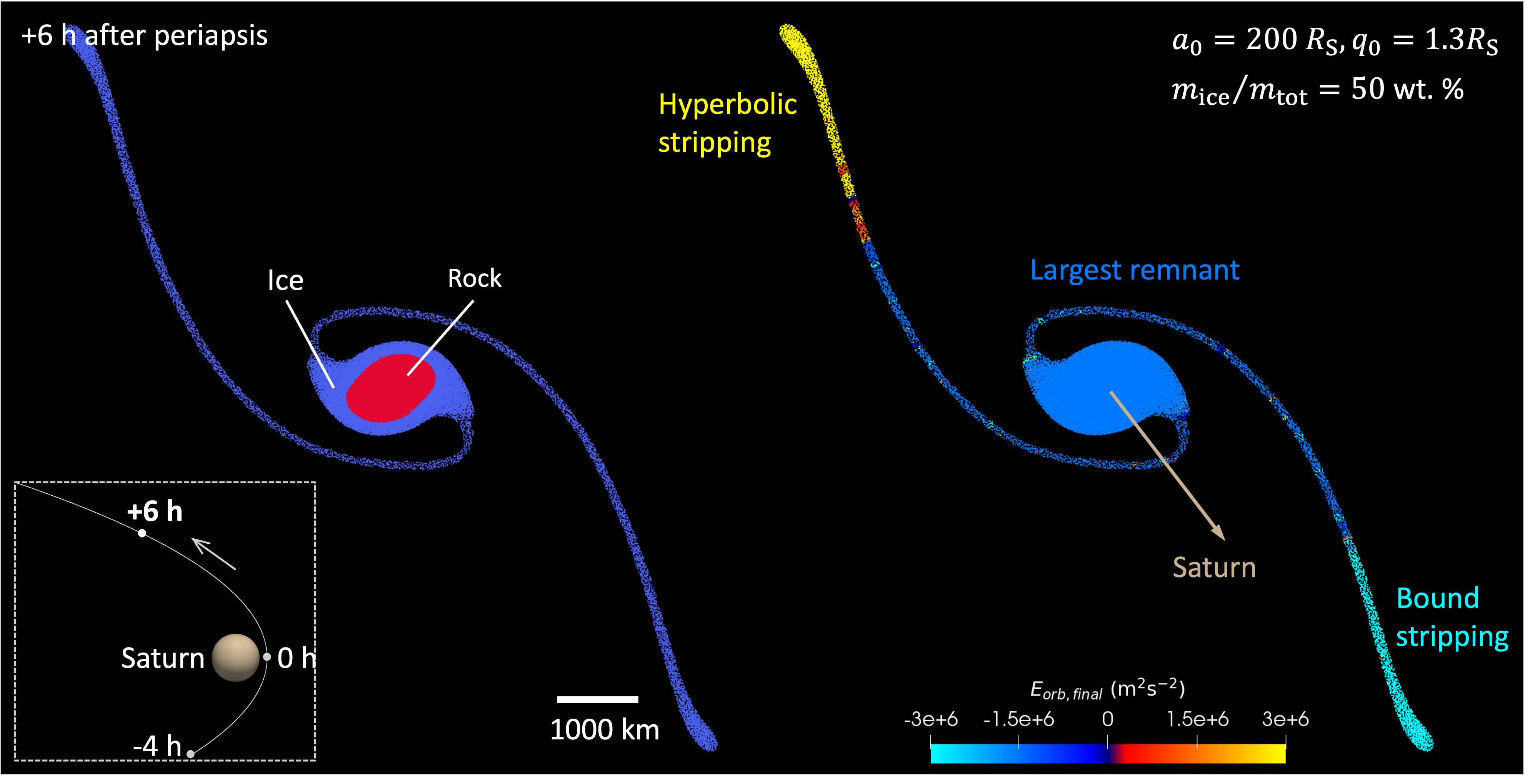}
\caption{SPH simulation of tidal stripping of Chrysalis during a close encounter with Saturn (case d5 in Table~\ref{tab:simu_setup}). The initial body is assumed to be differentiated with 50~wt.\% ice and 50~wt.\% rock. The pre-encounter semi-major axis is $200R_{\rm S}$, with a periapsis of $1.3R_{\rm S}$. The snapshot shows a cross section 6~h after periapsis. In the left panel, red particles indicate the rock core and blue the ice mantle. The full simulation spans 50~h. The right panel shows the specific orbit energy of each particle at the end, where a positive value means a hyperbolic orbit and negative for bound orbit. For reference, the pre-encounter orbital energy is about $-1.5\times10^6$~m$^2$s$^{-2}$. A supplementary video of this case is available at \url{https://github.com/jiaoyf-thu/tidal-stripping}.}
\label{fig:snapshot}
\end{figure*}

\begin{figure}[htb!]
\centering
\includegraphics[width=\linewidth]{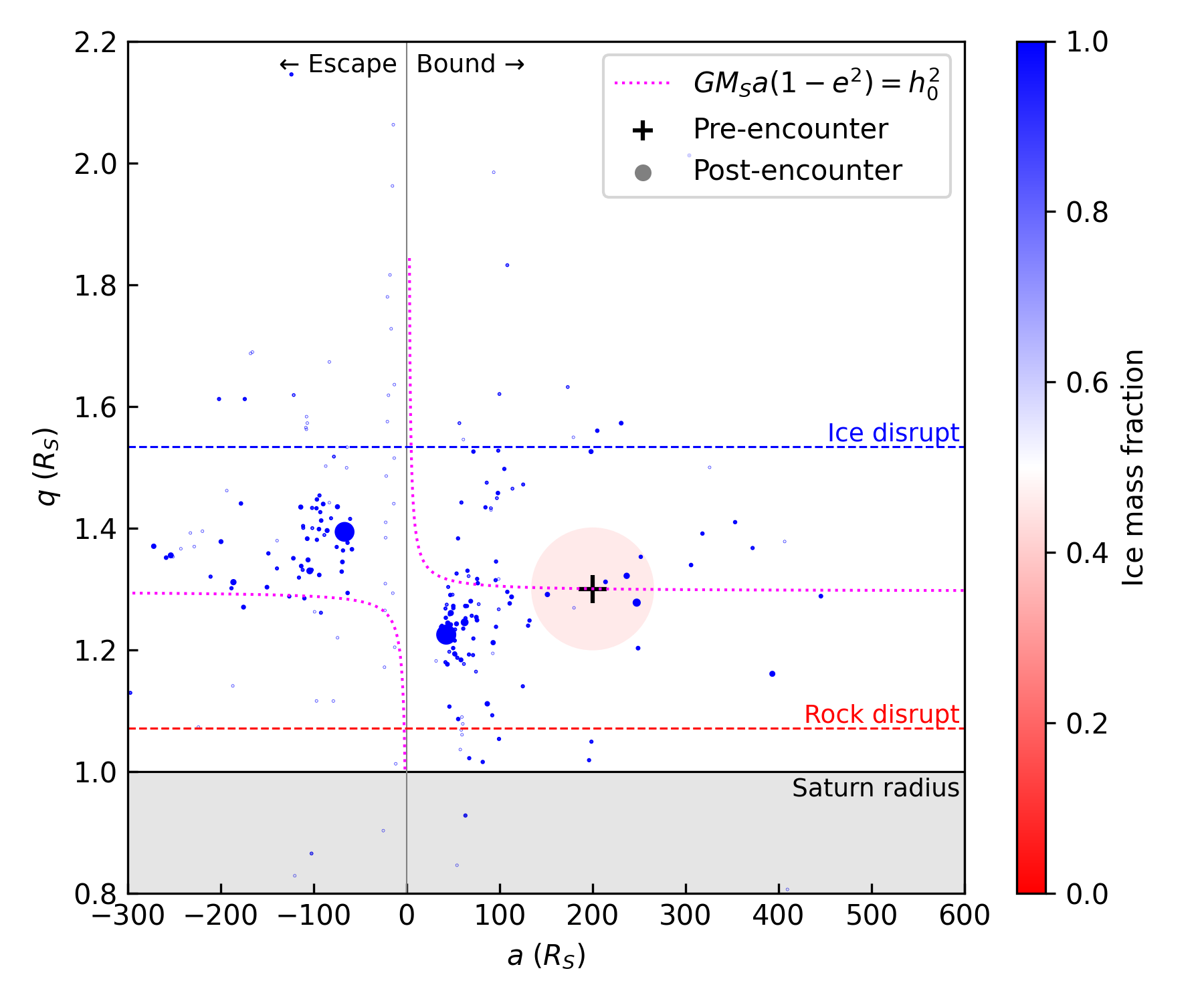}
\caption{Saturn-centric orbit distribution of the post-encounter Chrysalis and its debris at the end of SPH simulation (case d5 in Table~\ref{tab:simu_setup}). Each dot represents a cluster of particles, with its size representing the cluster mass and color for the ice fraction. The large light-red dot is the largest remnant of Chrysalis retaining the rock core, which experiences minor changes to the pre-encounter orbit (marked plus). The magenta dotted line indicates the specific angular momentum of the pre-encounter orbit of Chrysalis.}
\label{fig:ice-frac}
\end{figure}

As a typical tidal stripping result, Figure~\ref{fig:snapshot} presents a snapshot of Chrysalis during its close encounter with Saturn (case d5 in Table~\ref{tab:simu_setup}).
In this case, Chrysalis has a periapsis of $q_0=1.3R_{\rm S}$ that lies between the disruption limits for ice and rock, where tidal forces are strong enough to strip off the icy mantle but insufficient to disrupt the rocky core---similar to \cite{canup2010origin}.
Most of the rock particles are retained in the largest remnant of Chrysalis.
The stripped ice particles separate into two branches: one gains more orbital energy and becomes hyperbolic, while the other loses orbital energy, i.e., becomes bound to Saturn, migrates to smaller semi-major axis, and may ultimately form the rings.
Roughly, these two branches are from the farside and nearside of Chrysalis at its periapsis, according to the orbital energy distribution in \cite{dones1991recent}.
At the end of each SPH simulation, we identify all the debris clusters using a density-based spatial clustering algorithm \citep{jiao2024sph}, and compute their Saturn-centric orbits and compositions.
Figure~\ref{fig:ice-frac} shows the post-encounter orbits of these clusters. For reference, the specific orbit energy $E_{\rm orb}$ and the specific angular momentum $h$ can be derived from
\begin{subequations}
\begin{equation}
    E_{\rm orb} = -\frac{GM_{\rm S}}{2a},
\end{equation}
\begin{equation}
    h = \sqrt{GM_{\rm S}a(1-e^2)},
\end{equation}
\end{subequations}
where $G=6.6743\times10^{-11}$~m$^3$\,kg$^{-1}\,$s$^{-2}$ is the gravitational constant, $M_{\rm S}$ is the mass of Saturn. With $e\simeq1$, we have $h\simeq \sqrt{2GM_{\rm S}q}$.
Hyperbolic clusters have a negative semi-major axis ($a_{\rm h}<0$) and therefore a positive orbital energy, indicating that they are unbound. In contrast, the bound clusters have $0<a_{\rm b}\lesssim a_0$, remaining gravitationally bound to Saturn but with a more negative orbital energy than the pre-encounter value $E_{\rm orb,0}$.
Note that both the bound and unbound materials are predominately ice, while the largest remnant is rock-rich.
For clarity, we use the subscript ``h'' to represent the hyperbolic branch, and ``b'' for the bound branch.
This holds true for most of the cases except case~i18 (80~wt.\% ice, $q_0=1.5R_{\rm S}$), where tidal forces are relatively weak and both branches are still bound to Saturn but with different semi-major axes (see Figure~\ref{fig:snapshot-all}).

For the specific angular momentum, the hyperbolic and bound branches lie on opposite sides of the pre-encounter periapsis $q_0$, so that their angular momenta are respectively higher and lower than the pre-encounter value $h_0$.
As a consequence, the orbit of the largest remnant remains nearly unchanged relative to the pre-encounter orbit.
Recall that the largest remnant is ultimately removed either by colliding with Saturn or by becoming hyperbolic, as described in Section~\ref{sec2.2}.
However, the spin rate and bulk density of the largest remnant have been changed after tidal stripping, which could be important during multiple encounters (before final loss) and will be addressed in Section~\ref{sec3.3}.

\begin{figure}[htb!]
\centering
\includegraphics[width=\linewidth]{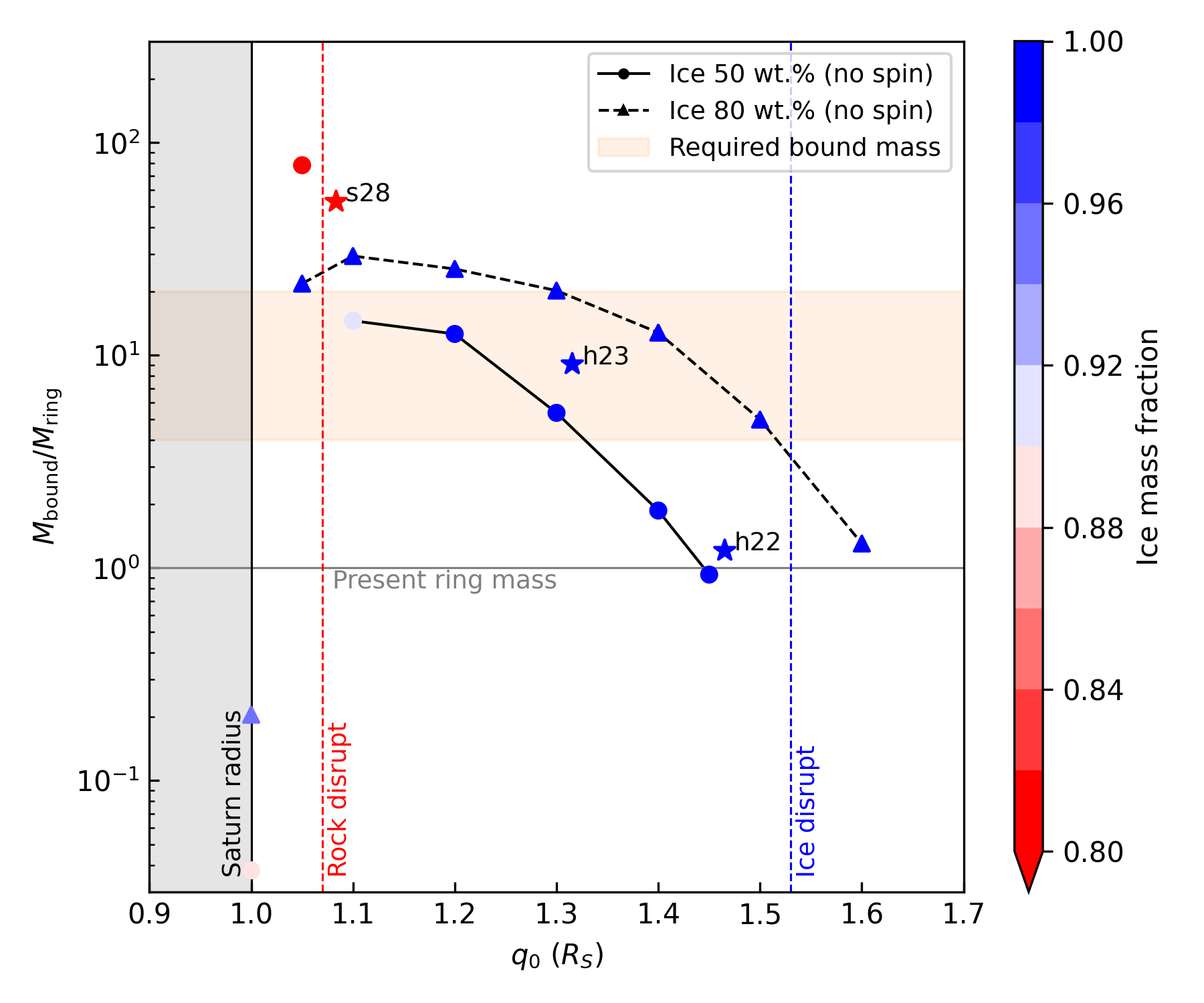}
\caption{The bound mass from tidal stripping as a function of periapsis $q_0$, compared with the present ring mass \citep{iess2019measurement} and the required bound mass (Section~\ref{sec3.4}). Solid and dashed lines represent the results of different ice-to-rock ratios, with the largest remnant excluded from the bound mass. Isolated circular/triangular markers indicate core-disrupted or intersecting-Saturn cases. Star markers labeled with case numbers have been corrected to the equivalent periapsis at no spin (see text). The colors represent the ice mass ratio of the bound mass, with blue for pure ice.}
\label{fig:ring-ice}
\end{figure}

Regarding the bound branch of tidally stripped debris, Figure~\ref{fig:ring-ice} presents their total mass (excluding the largest remnant) and composition, as a function of the periapses and ice-to-rock ratios of Chrysalis.
Generally, the bound mass increases with decreasing $q_0$ due to stronger tidal forces, while the material remains nearly pure ice (blue points).
With a higher ice fraction, the lower bulk density makes it easier to strip material from Chrysalis, thus producing more bound mass.
For tidal stripping at small $q_0$ (before core disruption occurs), some of the stripped mass may have $q_{\rm b}<R_{\rm S}$ to impact Saturn, which has been removed from the bound mass.
The upper limit of the bound mass is expected to be less than half of the total ice-mantle mass, since the remaining half would escape on hyperbolic trajectories.
In our simulations, the maximum bound mass occurs at $q_0=1.1R_{\rm S}$. At this location, the bound material contains 11\% of the initial Chrysalis mass for an initial ice fraction of 50~wt.\%, and 23\% of Chrysalis for an initial ice fraction of 80~wt.\%. These values correspond to 14 and 29 times the present ring mass (adopting the mean value of $M_{\rm ring}=1.54\times10^{19}$~kg).
In case d5, with $q_0=1.3R_{\rm S}$ and 50~wt.\% ice, the bound mass is 5.4 times the present ring mass, in close agreement with the 6.5$\times$ value from the high-resolution case d6, and both cases have bound material of 100\% ice.
Even at the ice disruption limit, where tidal forces are relatively weak, we can still expect a bound mass comparable to that of the current rings. This estimate is statistically robust, as it is derived from $\sim 10^3$ SPH particles out of a total of $10^5$.
Most of the tidal stripping cases produce bound material of $>98$~wt.\% ice (see Table~\ref{tab:simu_setup}).
Therefore, provided a periapsis occurs between the ice and rock disruption limits, the preferential tidal stripping of the ice mantle from Chrysalis can produce debris with both mass and composition resembling the present rings.

\subsection{Multiple tidal encounters}\label{sec3.3}

\begin{figure}[tb!]
\centering
\includegraphics[width=\linewidth]{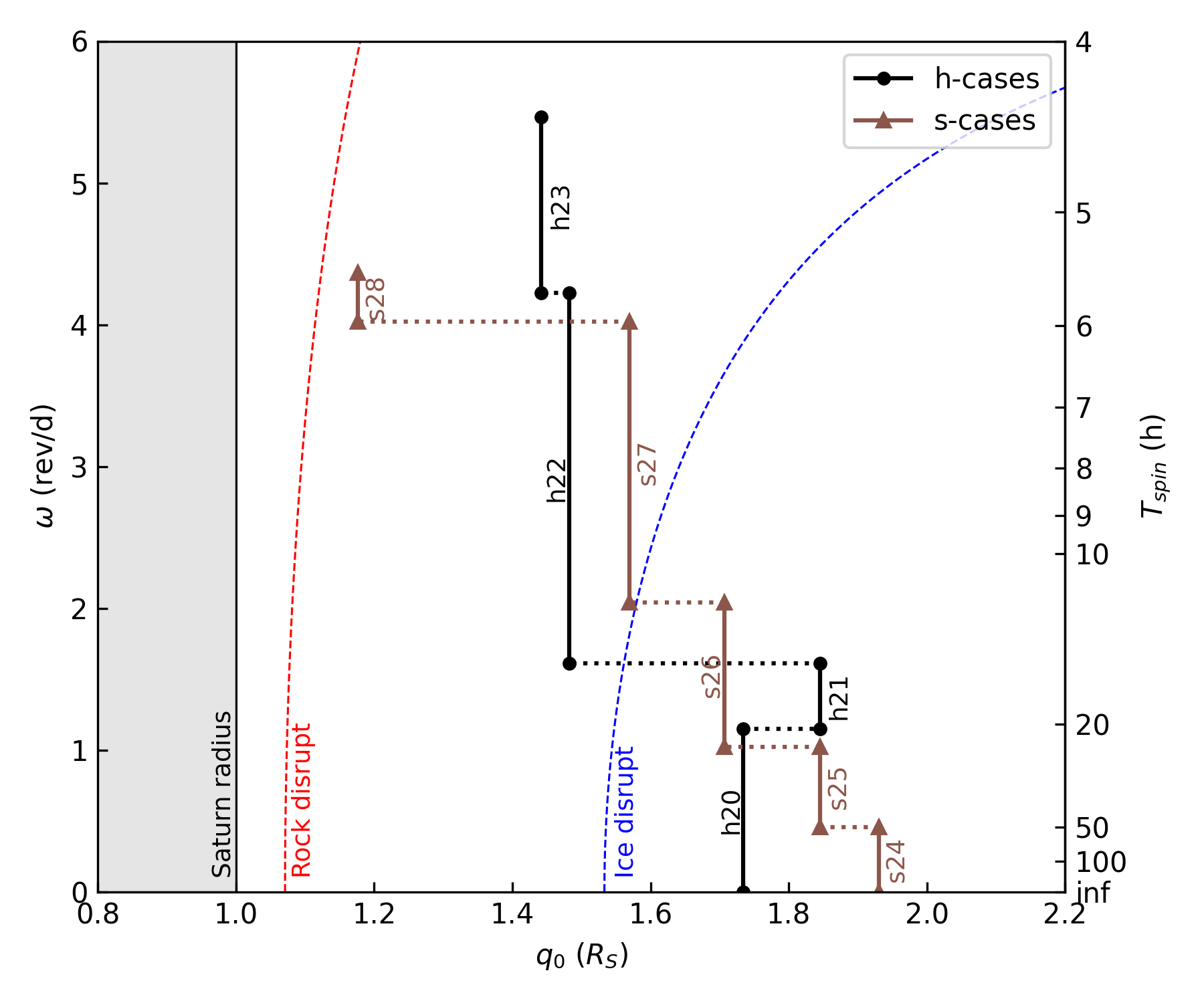}
\caption{Spin evolution during multiple close encounters of Chrysalis. The h- and s-cases correspond to those presented in Table~\ref{tab:simu_setup} and Figure~\ref{fig:para-orbit}~(b--c). In each case, the solid line represents a close encounter SPH simulation where $q_0$ is constant and the spin rate $\omega$ changes, while the dotted line assumes that the spin rate doesn't change between two close encounters. Both cases start with zero spin and 50~wt.\% ice. The parabolic Roche limits have been corrected with a spin-dependent factor \citep{malamud2020tidal}.}
\label{fig:multi-encounter}
\end{figure}

During close encounters, as shown in Figure~\ref{fig:ice-frac}, the largest remnant retains its pre-encounter orbit. However, it may be spun up by tidal torques due to the phase lag of deformation.
The centrifugal forces on a spinning body would then make it easier to be tidally disrupted in the next encounter. For simplification, here we introduce a spin-dependent factor $f_{\rm spin}$ to correct the Roche limit \citep{malamud2020tidal}
\begin{equation}
    f_{\rm spin} = \left (1 - \frac{\omega^2}{\omega^2_{\rm crit}} \right )^{-\frac{1}{3}},
\end{equation}
where $\omega_{\rm crit}=\sqrt{\frac{4}{3}\pi G\rho_0}$ is the critical spin rate for a strengthless spinning body with density of $\rho_0$. Note that $\omega<\omega_{\rm crit}$, and thus $f_{\rm spin}>1$: the Roche limit is increased for a spinning body.
Our h- and s-cases simulate multiple encounters of Chrysalis shown in Figure~\ref{fig:para-orbit}~(b--c).
Figure~\ref{fig:multi-encounter} shows that the spin rate of Chrysalis could have been gradually increased during multiple encounters, up to $\sim5$~rev\,d$^{-1}$, in which case the ice Roche limit is extended by a factor of 1.3 compared to the no-spin case.
We also note that a smaller $q_0$ leads to a larger spin change, e.g., cases h22 and s27, since the tidal deformation and the resulting torques are more effective. However, the spin cannot grow indefinitely: the stripped mass would carry away some of the body's angular momentum before it reaches the critical spin rate, e.g., cases h23 and s28.

We could define an equivalent periapsis at no spin, $q_0^*=q_0/f_{\rm spin}$, where $q_0$ denotes the physical periapsis with spin. This suggests that, for a spinning body, the encounter result is equivalent to that with a lower periapsis at no spin and is therefore more tidally disruptive.
For cases h22, h23 and s28, where tidal stripping or disruption occurs, we plot their equivalent periapses and the bound mass in Figure~\ref{fig:ring-ice}, and find that they are comparable to our previous no-spin results with an initial 50~wt.\% ice.
In particular, case s28 has $q_0=1.18R_{\rm S}$ but an equivalent value of $q_0^*=1.08R_{\rm S}$, and thus experiences a core disruption.

In addition to the spin effects, tidal stripping would remove the low-density ice material from the largest remnant, reducing its mass while increasing its bulk density.
In extreme cases, e.g., case i13 with $q_0=1.05R_{\rm S}$ and 80~wt.\% ice, the remaining core has only $\sim$52\% of its pre-encounter mass and a decreased ice fraction of $\sim$62~wt.\%.
A higher bulk density would then make it more resistant to tidal forces during next encounter, which may partially cancel the spin effects but requires further investigations.
Also note that the orbital inclination is not considered in our simulations, which may change during multiple encounters (see Figure~\ref{fig:para-orbit}).
This would have caused a misalignment between the tidal torques and the spin vector, whereas our simulations assume perfect alignment.
Therefore, our spin evolution results should be interpreted as an upper limit.

\subsection{Dynamical evolution of the bound debris}\label{sec3.4}

The bound debris from tidal stripping still has highly eccentric and possibly inclined orbits, while the present rings are circular and equatorial.
As suggested by \cite{hyodo2017ring}, the apsidal and nodal precessions driven by Saturn's $J_2$ term would disperse the debris orbits into a torus-like structure in hundreds of years.
Mutual collisions between the debris particles would then effectively damp their eccentricities and inclinations to form the present rings \citep{morbidelli2012explaining}.
The debris can further break up into small fragments during these catastrophic collisions.
The collisional timescale is typically a few thousand years \citep{hyodo2017ring}, depending on the number density of particles.
Regarding the orbital angular momentum concern rasied by \cite{crida2025age}, assuming that the vertical angular momentum of the bound debris is conserved, the equivalent semi-major axis of the damped rings is then
\begin{equation}
    a_{\rm ring} = a_{\rm b} (1-e_{\rm b}^2) \cos^2 i_0 \simeq 2q_{\rm b} \cos^2 i_0,
\end{equation}
which is determined by both the initial periapsis $q_{\rm b}$ and initial inclination $i_0$ of the bound debris.
To match the present ring orbits with $a_{\rm ring}<2.5R_{\rm S}$, the initial periapsis must satisfy  $q_{\rm b}<1.25R_{\rm S}$ when $i_0=0^\circ$. Recalling that $q_{\rm b}\lesssim q_0$, the pre-encounter periapsis $q_0$ of Chrysalis could be slightly larger, such as $q_0=1.3R_{\rm S}$ as shown in Figure~\ref{fig:ice-frac}.
On the other hand, the actual inclination $i_0$ may range from $0^\circ$ to $90^\circ$ (Figure~\ref{fig:para-orbit}). Adopting $i_0=45^\circ$ yields $q_{\rm b} \simeq a_{\rm ring}<2.5R_{\rm S}$, which further relaxes the requirement for a very close periapsis encounter suggested by \cite{crida2025age}.

We note that the bound debris is subject to perturbations from Saturn's large moons, such as Titan, which can remove as much as $\eta_1=$~40\%--70\% of the total mass in the first few thousand years \citep{hyodo2017ring}.
The mass of the rings also gradually decreases due to viscous spreading \citep{crida2019saturn} and micrometeoroid-driven processes such as ballistic transportation and mass loading \citep{estrada2023constraints}.
It has been estimated that the initial mass of Saturn's rings would have been between 1--3 Mimas masses if they formed a few 100~Myr ago \citep{crida2025age}, implying a long-term mass loss of $\eta_2=$~60\%--85\%.
Therefore, the final mass of the rings is expected to be $(1-\eta_1)(1-\eta_2)$ of the bound mass, roughly 5\%--25\%. This corresponds to an initial bound mass about 4 to 20 times the present rings, consistent with our simulation results as shown in Figure~\ref{fig:ring-ice}.

\section{Discussion}

This work has focused on testing Saturn's rings formation under the recent-lost-moon hypothesis by \cite{wisdom2022loss}. However, some questions remains unanswered and should be addressed in future investigations.
The largest remnant after tidal stripping is expected to either impact Saturn or become hyperbolic within less than few kyr after the close encounter(s) (see Figure~\ref{fig:para-orbit}).
If it impacted Saturn, which is the most likely case (80\%), the remnant would be completely lost. How would such a giant impact alter Saturn's atmosphere and/or interior, and could any observable signatures persist to the present day?
If the remnant escaped into a heliocentric orbit (20\%), it would be subject to strong planetary scattering, similar to that experienced by Centaurs, leading to dynamical instability on 1--10 Myr timescales \citep{di2020centaur}. Is it possible to detect the largest remnant and/or tidally stripped hyperbolic debris if they still exist?
A $\sim$1000~km large object with a young icy surface and rapid rotation would be a possible candidate.
Also, for the bound debris that ultimately forms the rings, would some of them---with their highly eccentric initial orbits---collide with the other moons as a transient planetocentric impactor population?
Such debris would be expected to produce relatively low velocity, oblique impacts on the inner moons, which may be consistent with the elliptical craters identified on the surfaces of Tethys and Dione \citep{ferguson2022unique}.
In addition, some of Saturn's irregular distant satellites are found to have a very steep size-frequency distribution \citep{ashton2021evidence,ashton2025retrograde}, suggesting a recent collisional disruption history.
Some of the stripped debris from Chrysalis could reach an apoapsis distance of several hundred $R_{\rm S}$, where it could have disrupted a pre-existing irregular satellite.

\section{Summary}

The recent-lost-moon hypothesis \citep{wisdom2022loss} is one of the more promising scenarios to explain the distinctive Saturnian system, particularly the origin of its young, icy rings.
The pre-existing moon, Chrysalis, is constrained to be a $\sim M_{\rm Iapetus}$ object on a highly eccentric orbit before its close encounter with Saturn.
Our smoothed particle hydrodynamics simulations demonstrate that preferential tidal stripping of the ice mantle from a differentiated Chrysalis can produce debris that will ultimately form a ring with both mass and composition resembling the present rings, supporting the recent-lost-moon hypothesis.
The required periapsis distance $q_0$ lies between the parabolic Roche limits for ice and rock, approximately 1.07$R_{\rm S}$ to 1.53$R_{\rm S}$.
Moreover, multiple close encounters can extend the effective disruption limit by spinning up the body, enhancing the tidal stripping efficiency.
Following these close encounters, the rocky remnant of Chrysalis would have been removed in less than few kyr, either by collision with Saturn or ejection onto a hyperbolic orbit.
The tidally stripped debris would ultimately evolve into the present rings through long-term dynamical processes, as described in \cite{hyodo2017ring}.




\begin{contribution}

Y.J. performed the SPH simulations, analyzed the numerical results, and led the writing of the manuscript.
F.N. initiated the conception of this project and contributed to the interpretation of the results.
J.W. and R.D. contributed to defining the parameter space and interpretation of the results. 
All authors contributed to the revision of the manuscript.


\end{contribution}


\appendix
\twocolumngrid
\setcounter{figure}{0}
\renewcommand{\thefigure}{A\arabic{figure}}

\section{SPH Code validation}

The tidal encounter simulations are performed with the smoothed particle hydrodynamics code developed by \citet{jiao2024sph,jiao2024asteroid}, which is publicly available at \url{https://sphsol-tutorial.readthedocs.io}. The code was originally designed for modeling hypervelocity impacts on planetary bodies, and is extensible to include additional particle-level forces.
To model the close-encounter process, we adopt a co-moving frame centered on the passing object's center of mass, with the updated momentum equation as
\begin{equation}
    \frac{{\rm d}\bm{v}}{{\rm d}t} = \frac{1}{\rho}\left(\nabla\cdot\bm{\sigma}\right) + \bm{a}_{\rm grav} + \bm{a}_{\rm tide},
\end{equation}
where $\rho$ is the particle density, $\bm{\sigma}$ is the stress tensor, $\bm{a}_{\rm grav}$ is the gravitational attractions from all the other particles, and $\bm{a}_{\rm tide}$ is the differential force of the planet's gravity, i.e., the tidal force.
The tidal force term is then incorporated using a first-order Taylor expansion, assuming that the passing object is small compared to the closest approach distance
\begin{equation}
      \bm{a}_{i, \rm tide} = \frac{GM_{\rm pla}}{r_{\rm com}^3} (3\hat{\bm{r}}_{\rm com}\hat{\bm{r}}_{\rm com}^\top - \bm{I}) \Delta \bm{r}_i,
      \label{eq:tide}
\end{equation}
where $M_{\rm pla}$ is the planet mass, the passing object's trajectory $\bm{r}_{\rm com}$ is obtained by interpolating a pre-tabulated dataset, $\hat{\bm{r}}_{\rm com}$ is the unit vector in the direction of $\bm{r}_{\rm com}$, $\bm{I}$ is the identity matrix, and $\Delta\bm{r}_i$ is the position of particle $i$ relative to the passing object's center of mass.
To validate the modified code, we performed numerical simulations of two scenarios: (1) tidal disruption of comet Shoemaker--Levy 9 with Jupiter \citep{asphaug1994density}, and (2) tidal disruption of a Titan-mass passing comet with Saturn \citep{hyodo2017ring}.

\subsection{Tidal disruption of Shoemaker--Levy 9}

\begin{figure}[b!]
\centering
\includegraphics[width=\linewidth]{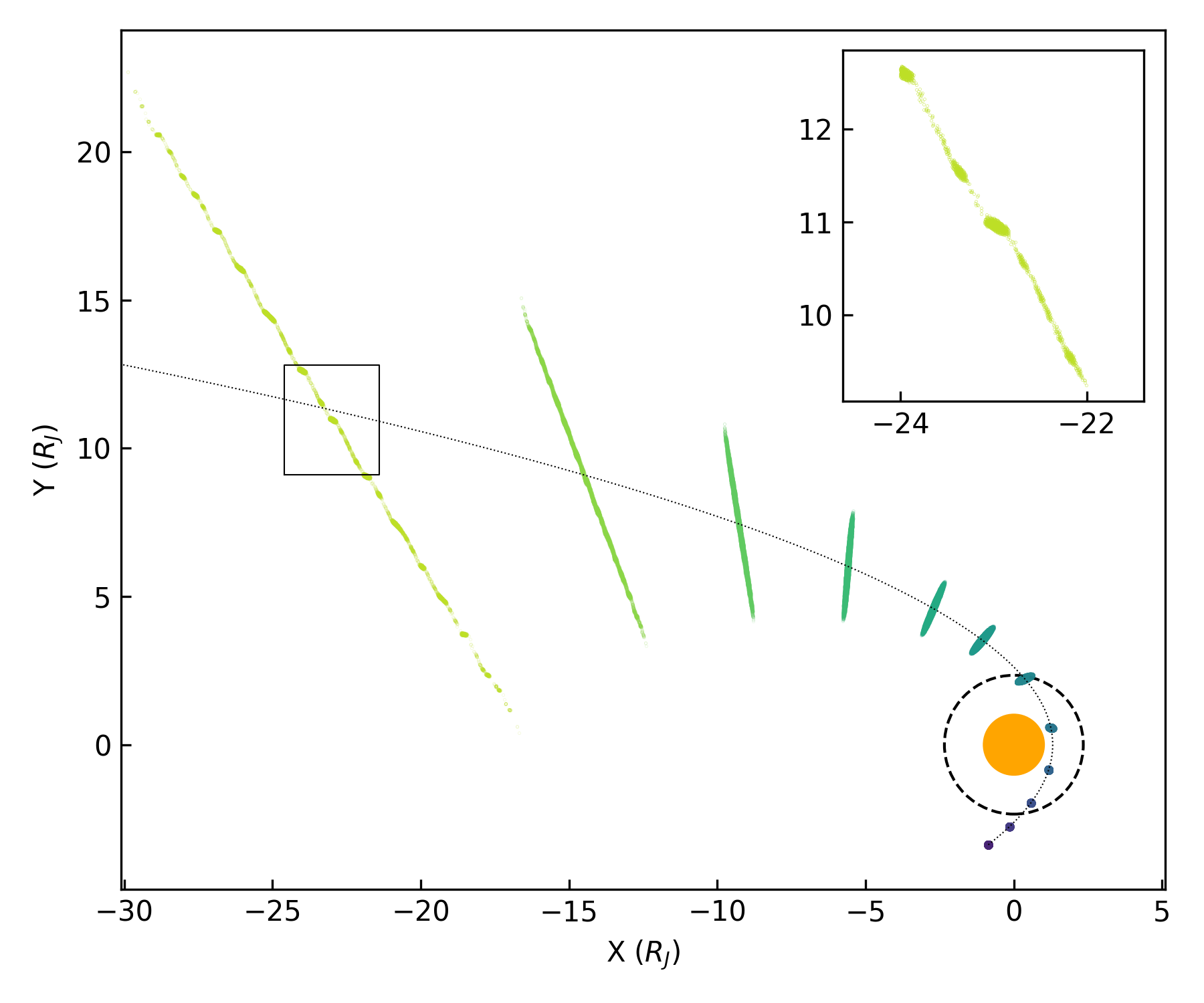}
\caption{Tidal disruption of SL9 from $-$2~h to $+$31~h relative to the periapsis. The comet debris are magnified by 10,000 times for visualization, with color indicating the time series. The inset shows a zoom-in view of the $+$31~h results, where self-gravitational clumps form. The orange circle represents Jupiter, and the dashed circle represents the disruption limit for parabolic, viscous bodies \citep{sridhar1992tidal}.}
\label{fig:sl9-simu}
\end{figure}

\begin{figure}[tb!]
\centering
\includegraphics[width=\linewidth]{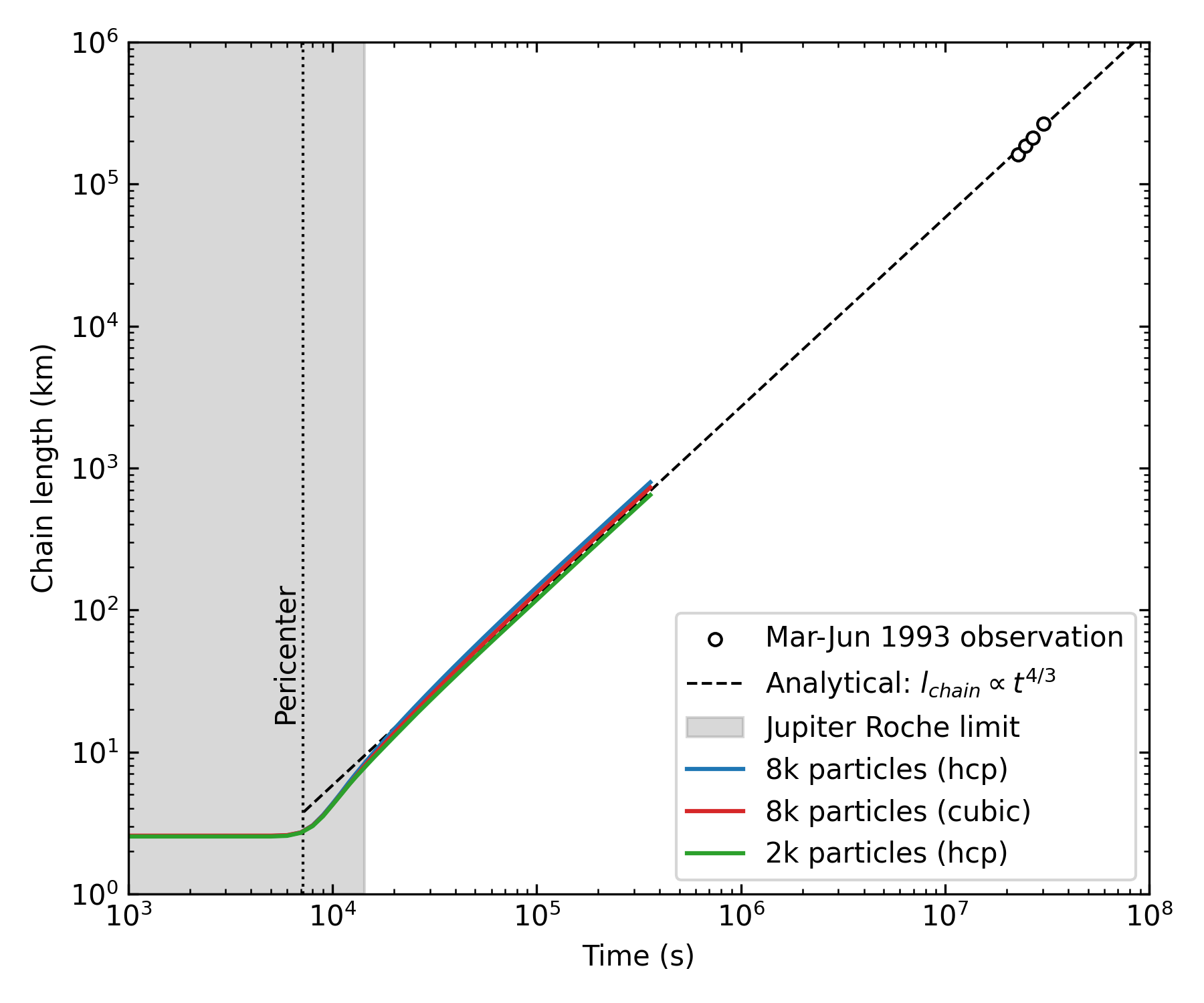}
\caption{SL9 chain length over time in our simulations. Different colored lines represent different particle resolutions or packing methods, compared with analytical models \citep{sridhar1992tidal} and observations \citep{asphaug1994density}.}
\label{fig:sl9-chain}
\end{figure}

As a typical tidal disruption scenario, the comet Shoemaker--Levy 9 (SL9) experienced a close encounter with Jupiter in 1992, and broke apart into $\sim$21 pieces of comparable size.
Here we use our code to reproduce the numerical simulations of SL9's tidal disruption in \citet{asphaug1994density}, which employed a strengthless parent body with diameter of 1.5~km and bulk density of 0.5~g\,cm$^{-3}$. The Tillotson equation of state for ice \citep{melosh1989impact} is adopted for the comet material. The trajectory is assumed to be parabolic with periapsis of $1.3R_{\rm J}$, which is below the disruption limit of $\sim2.3R_{\rm J}$ \citep{sridhar1992tidal}.
Our simulation starts from 2 hours before the pericenter (outside the Roche limit) and lasts for 100 hours. As shown in Figure~\ref{fig:sl9-simu}, SL9 becomes elongated and breaks apart into a debris chain after its Jupiter encounter.
The debris then separate into clumps under self-gravitational instabilities.
Figure~\ref{fig:sl9-chain} shows that the chain length increases with time as $l_{\rm chain} \propto t^{4/3}$ as predicted by analytical models \citep{sridhar1992tidal}, which also matches the observed lengths in 1993 \citep{asphaug1994density}.
For comparison, we also perform simulations with different particle resolutions (2,000 and 8,000 particles) and packing methods (hexagonal close packing and simple cubic), all of which exhibit good conservation of results.

\subsection{Tidal disruption of a Titan-mass comet}

\begin{figure}[htb!]
\centering
\includegraphics[width=\linewidth]{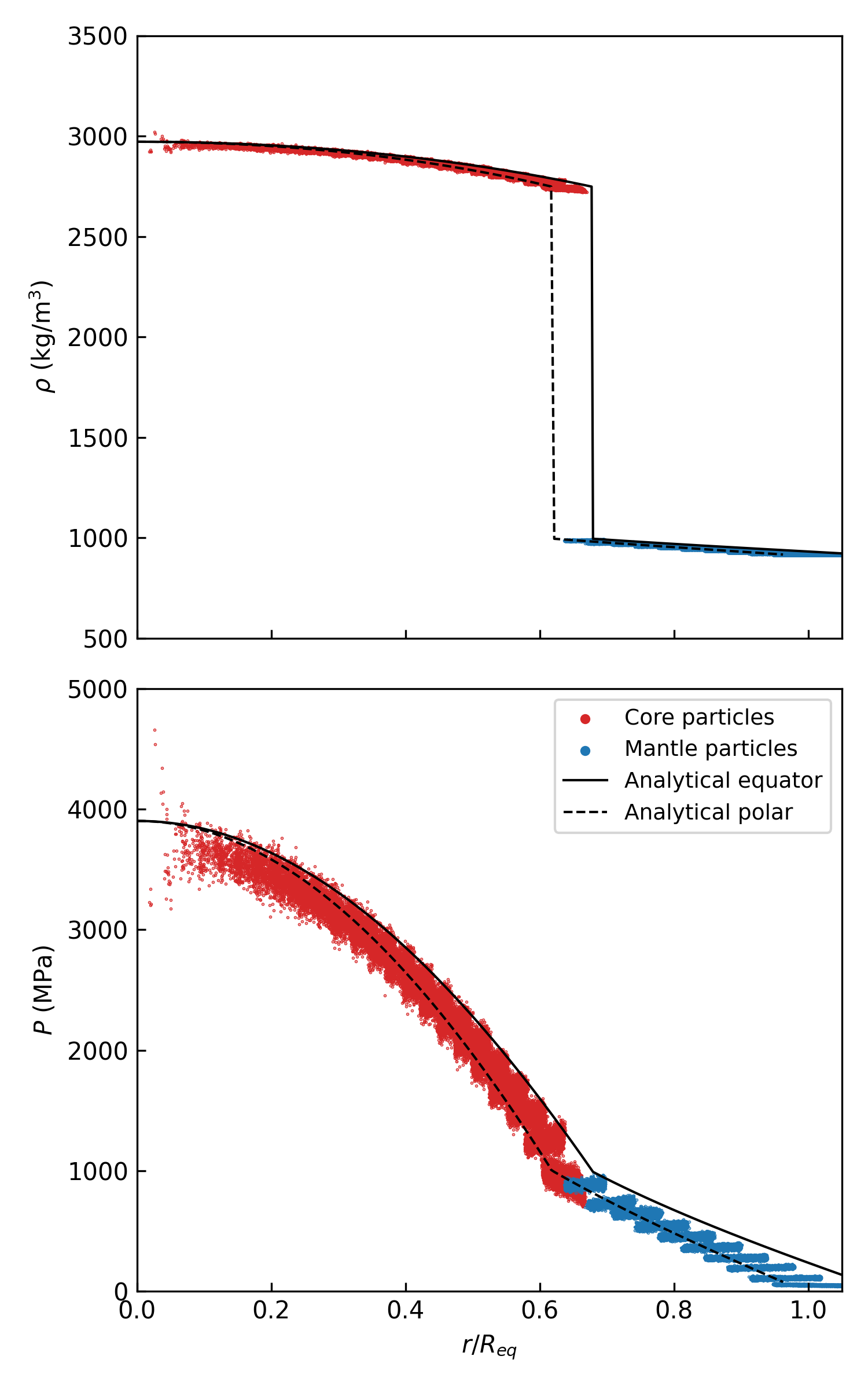}
\caption{Initial profile for a differentiated object under self gravity and with spin period of 8~h. The analytical model are from \citet{ruiz2021effect}. The red points indicate rock core particles, with blue for ice mantle. The particle configuration is then used as the initial conditions for the simulations shown in Figure~\ref{fig:hyodo-re1}.}
\label{fig:ini-profile}
\end{figure}

\begin{figure*}[htb!]
\centering
\includegraphics[width=\linewidth]{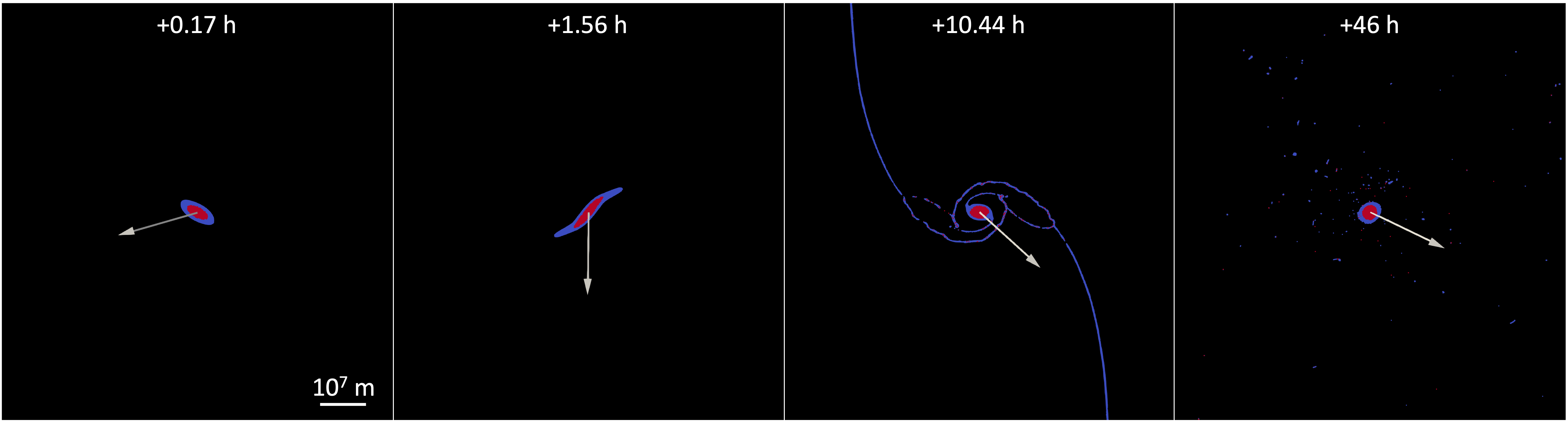}
\caption{Cross-section snapshots of tidal disruption of a 10$^{23}$~kg, differentiated body by Saturn, with initially no spin, periapsis of $7.0\times10^{7}$~m, velocity at infinity of 3~km\,s$^{-1}$. All snapshots are presented in the center-of-mass frame of the body, with the arrow pointing to Saturn. Timestamps are referenced to the periapsis. Red particles indicate the rock core, and blue particles indicate the ice mantle. The scenario reproduces the setup and results in \citet{hyodo2017ring} (see their Figure~10).}
\label{fig:hyodo-re2}
\end{figure*}

\begin{figure*}[htb!]
\centering
\includegraphics[width=\linewidth]{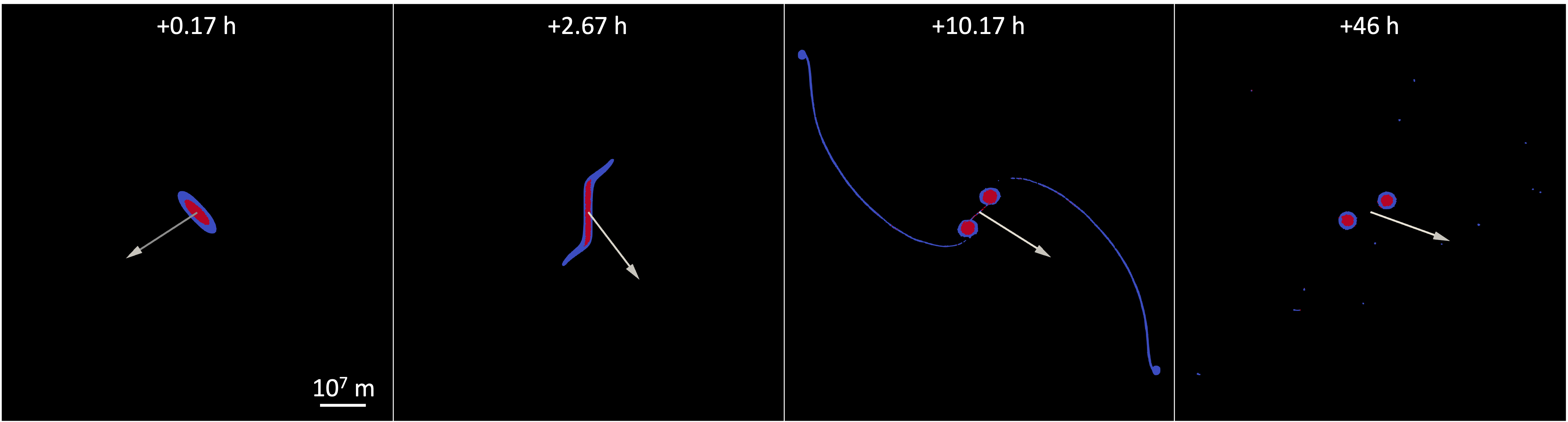}
\caption{Same as Figure~\ref{fig:hyodo-re2}, but for a body with initial retrograde spin period of 8~h, periapsis of $4.2\times10^{7}$~m, velocity at infinity of 3~km\,s$^{-1}$. The scenario reproduces the setup and results in \citet{hyodo2017ring} (see their Figure~11).}
\label{fig:hyodo-re1}
\end{figure*}

Since we are considering large objects of $\sim$1000~km in this work, it is necessary to begin with a state in hydrodynamic equilibrium under the body's self gravity and rotation.
To address this, we use the open-source WoMa code \citep{ruiz2021effect} to create the density and pressure profiles analytically, and map the profile into particle configuration. The generated particles are then relaxed in our SPH code as described in \citet{vsevevcek2019impacts}.
Figure~\ref{fig:ini-profile} shows an example of a two-layer structure consisting of an ice mantle and a rock core, where the relaxed particles remain closely consistent with the analytical models, and are numerically stable at the start of simulations.
With the initial hydrodynamic equilibrium, we simulate the tidal disruption of a Titan-mass ($10^{23}$~kg) comet by Saturn, with parameters adopted from \citet{hyodo2017ring}.
The body is assumed to be differentiated with an ice-to-rock mass ratio of 50:50, and modeled with the Tillotson equation of state for water ice and basalt for each component \citep{melosh1989impact}. The total number of SPH particles is $10^5$, same as \citet{hyodo2017ring}.
The comet orbit is assumed to be hyperbolic with velocity at infinity of 3~km\,s$^{-1}$.
Figures~\ref{fig:hyodo-re2}--\ref{fig:hyodo-re1} show two scenarios with different periapsis and initial spins.
The results successfully reproduce the tidal deformation and destruction process of a differentiated body, with two main branches of spiral arms that is mostly ice, as presented in \citet{hyodo2017ring}.
In Figure~\ref{fig:hyodo-re1}, the body is split into two large objects due to the strong tidal forces from Saturn.
The capture efficiency is 0.0899 and 0.0699 in our two simulations, which also matches the results in \citet{hyodo2017ring}.
These results demonstrate the capability of our code in modeling the tidal disruption process.



\newpage
\bibliography{sample701}{}
\bibliographystyle{aasjournalv7}



\end{document}